\documentclass[review]{elsarticle}
\usepackage{url,hyperref,microtype,subcaption}
\usepackage{hyperref}
\usepackage{amsmath}

\journal{Journal of \LaTeX\ Templates}

\begin{document}

\begin{frontmatter}

\title{Inferring the connectivity of coupled oscillators and anticipating their transition to synchrony through lag-time analysis}

\author{Inmaculada Leyva\,$^{1,2}$ and Cristina Masoller\,$^{3*}$}
\address{$^{1}$ Complex Systems Group \& GISC, Universidad Rey Juan Carlos, Madrid, Spain \\ $^{2}$Center for Biomedical Technology, Universidad Polit\'ecnica de Madrid, Madrid, Spain \\$^{3}$Departament de Fisica, Universitat Politecnica de Catalunya, Barcelona, Spain  }

\begin{abstract}
The synchronization phenomenon is ubiquitous in nature. In ensembles of coupled oscillators, explosive synchronization is a particular type of transition to phase synchrony that is first-order as the coupling strength increases. Explosive sychronization has been observed in several natural systems, and recent evidence suggests that it might also occur in the brain. A natural system to study this phenomenon is the Kuramoto model that describes an ensemble of coupled phase oscillators. Here we calculate bi-variate similarity measures (the cross-correlation, $\rho_{ij}$, and the phase locking value, PLV$_{ij}$)  between the phases, $\phi_i(t)$ and $\phi_j(t)$, of pairs of oscillators and determine the lag time between them as the time-shift, $\tau_{ij}$, which gives maximum similarity (i.e., the maximum of $\rho_{ij}(\tau)$ or PLV$_{ij}(\tau)$). We find that, as the transition to synchrony is approached, changes in the distribution of lag times provide an earlier warning of the  synchronization transition (either gradual or explosive). The analysis of experimental data, recorded from Rossler-like electronic chaotic oscillators, suggests that these findings are not limited to phase oscillators, as the lag times display qualitatively similar behavior with increasing coupling strength, as in the Kuramoto oscillators. We also analyze the statistical relationship between the lag times between pairs of oscillators and the existence of a direct connection between them. We find that depending on the strength of the coupling, the lags can be informative of the network connectivity.  
\end{abstract}


\end{frontmatter}

\section{Introduction}

Inferring the underlying connectivity of a complex system from the observed output signals is an important problem in nonlinear science with applications across disciplines. For example, functional brain networks are generated using statistical similarity measures applied to electroencephalogram (EEG) signals. In this approach, pairs of brain regions are considered functionally linked if there is high similarity between the time series recorded in the regions \cite{Rubinov2010,Bullmore2012}. However, the ability of the functional brain networks methodology to infer the underling structure is still an open problem \cite{bassett}. To demonstrate the potential of any approach as a clinical diagnostic tool it is necessary to test the inference methodology over datasets where the underlying connectivity is known. Many studies have been done in this direction, specially in gene networks \cite{Smith2002, Marbach2012}. However, network inference remains an open problem in the general case of dynamic units, where the interaction of the coupling structure and the dynamics of the units can hinder the detailed inference of the network connectivity (at the level single link).

In parallel to the inference problem, in the analysis of functional networks is specially important to be able to detect and to predict changes in the structure and in the dynamics of the system, which can reveal the proximity to a functional transition, such as variations on the  synchronization levels between brain areas. The sudden and nearly unannounced appearance of an epileptic crisis \cite{klaus} is a good example, where it is a challenge to obtain information from the often subtle changes in the system dynamics \cite{Van2013}. In some cases the system presents a very low synchronization level just before reaching full synchronization in a sudden, irreversible way.  This phenomenon, known as {\it explosive synchronization} (ES), has been observed in experimental systems \cite{Leyva2012, Boccaletti2016}, and it has been very recently hinted in the anesthetic-induced transition to unconsciousness \cite{Kim2016, Kim2017}, epilepsy \cite{Wang2017} or chronic pain \cite{Lee2018}. 

In this work we analyze numerical and experimental datasets obtained from Kuramoto oscillators \cite{Kuramoto1984} and from R\"ossler-like electronic oscillators \cite{datos28osc}, respectively.  We show how, taking into account the optimal lags in the computation of the similarity measure between each pair of time series, the resulting functional network allows to predict the proximity of the transition to synchrony, even when the early signs of the development of the process  are hidden, as happens in the case of explosive synchronization.  

\section{Datasets}

\subsection{Kuramoto oscillators}
We consider a network of $N$ Kuramoto phase oscillators, which is described by \cite{Kuramoto1984}:
\begin{equation}
\dot{\phi_i}=\omega_{i}+d \sum_{i=1}^{N} A_{ij} \sin \left(\phi_{j}-\phi_{i}\right)
\label{eq:kuramoto}
\end{equation}
where $\phi_i$ is the phase of the $i$th oscillator ($i$ = 1, …, $N$), $\omega_i$ is its associated natural frequency, drawn from the homogeneous frequency distribution $\omega$=[0,1]. The parameter $d$ is the coupling strength, and ${\bf A}=\{A_{ij}\}$ is the symmetric adjacency matrix: $A_{ij}=1$ if $i$ and $j$ are connected, else,  $A_{ij}=0$.  

We use an extra parameter that allows changing the nature of the synchronization transition from a soft, gradual transition to an abrupt, first-order-like transition. As explained in Ref. \cite{Leyva2013a}, this can be  done by explicitly imposing a {\it frequency dissasortativity} in the network, i. e., a constraint in the frequency differences between each pair of oscillators that are linked. With this procedure, we construct our network (i.e., the adjacency matrix) by randomly connecting the nodes up to obtain a preset average degree $\left < k \right >$ as in the usual Erd\"{o}s-Reyni model, but with the condition that a pair of oscillators  $i$ and $j$ can be connected only if  $\left| \omega_i – \omega_j \right| > \gamma$, with $\gamma$ a parameter that is refered to as {\it frequency gap}. This condition avoids that connected pairs of oscillators that have similar frequencies act as synchronization seeds. A consequence of this condition is that for high enough $\gamma$ the synchronization transition is explosive \cite{Leyva2013a}. 

We use this procedure to construct networks with $N$=50 oscillators, with mean degree $\left < k \right >$=5, and different values of the frequency gap $\gamma$.  Over each network, the dynamics described by Eq. \ref{eq:kuramoto} is simulated for a wide range of coupling values $d$. For each value of $d$, the time series of the phases of the 50 oscillators are analyzed.

\subsection{Electronic oscillators}

We also perform the network inference study over a set of experimental time series. The freely-available datasets, described in Ref. \cite{datos28osc}, come from an ensemble of $N$=28 networked R\"ossler chaotic electronic oscillators diffusely coupled through one of the variables. The structure of physical connections between oscillators (i.e., the adjacency matrix) is also provided, being a random matrix with 42 links (therefore 336 links do not exist). The dynamics of the variable describing the evolution of the each oscillator is given for a wide range of coupling strengths, capturing the transition from the unsynchronized behaviour to the synchronized one. As the oscillators are randomly coupled (i.e., there is no frustration parameter), the synchronization transition is not explosive. The time series have 30000 data points for every oscillator and coupling value.

\section{Methods}

\subsection{Bi-variate similarity measures}

The lagged Pearson coefficient, $\rho_{ij}(\tau)$, that is the absolute value of the cross-correlation coefficient, is used to quantify the similarity between the time series $x_i(t)$ and $x_j(t)$ in oscillators $i$ and $j$. When $x_i(t)$ and $x_j(t)$ with $t=1\dots T$ are each normalized to zero-mean and unit variance, $\rho_{ij}(\tau)$ can be calculated as
\begin{equation}
\rho_{ij}(\tau)=\frac{1}{T-\tau}\left|\sum_{t=1}^{T-\tau} x_i(t)x_j(t+\tau)\right|.
\label{eq:pearson}
\end{equation}
For each pair of oscillators $i$ and $j$ we calculate the lag, $\tau_{ij}$, that maximizes $\rho_{ij}(\tau)$. We search for the maximum in the interval $-\tau_{max}\le \tau \le \tau_{max}$ with $\tau_{max}=T/5$.  If several equal maximum values are found, the one with the smallest lag is selected. We measure the dynamical similarity between oscillators $i$ and $j$ as
\begin{equation}
S_{ij}=\rho_{ij}(\tau_{ij}).
\end{equation}

For the Kuramoto oscillators we calculate the cross-correlation between $\cos(\phi_i(t))$ and $\cos(\phi_j(t))$, while for the R\"ossler electronic oscillators, we calculate the cross-correlation between the experimental signals, i.e., the time series of the observed variables. The use of the cosine in the  Kuramoto oscillators is motivated by the fact that we can consider $\cos(\phi(t))$ an  ``observed'' variable (phases are usually not experimentally directly accessible, but they are calculated by using a suitable transformation, for example, the Hilbert transform).  On the other hand, this measure of similarity has the drawback that oscillators with similar  frequencies will have large cross-correlation values, regardless of the existence of a direct connection between them. 

To demonstrate that indeed the cross-correlation computed from the cosines of the phases contains useful information, we also quantify the dynamical similarity between Kuramoto oscillators $i$ and $j$ using the lagged phase locking value (PLV):
\begin{equation}
PLV_{ij}(\tau)=\frac{1}{T-\tau}\sum_{t=1}^{T-\tau}  e^{i (\phi_{i}(t)-\phi_{j}(t-\tau))}.
\label{eq:PLV}
\end{equation}
As with the cross-correlation, we search for the lag that maximizes $PLV_{ij}(\tau)$ in the interval $-\tau_{max}\le \tau \le \tau_{max}$ with $\tau_{max}=T/5$. 


\subsection{Global synchronization measures}

For the Kuramoto oscillators, the level of phase synchronization can be monitored by the order parameter given by: 
\begin{equation}
K=\left<\frac{1}{N}\left|\sum_{i=1}^{N} e^{i \phi_{i}(t)}\right| \right>_T
\label{eq:K}
\end{equation}
 where $\left<...\right>_T$ denotes a time average. In the general case, as the coupling strength $d$ increases, system represented by Eq.  \ref{eq:kuramoto} undergoes a soft, second-order like phase transition from an incoherent, desynchronized   ($K \sim$  0)  state to a synchronous ($K \sim$  1) state, where all oscillators ultimately acquire the same frequency.  

For the R\"ossler chaotic electronic oscillators we use, as a global measure of synchronization, the bivariate similarity measure, $S_{ij}$, averaged over all the network, $\left<S_{ij}\right>_{ij}$.

\subsection{Network inference}
Using time series recorded from a set of 12 randomly coupled R\"ossler chaotic oscillators \cite{Giulio} it has been recently reported \cite{Nico} that the mutual lags between connected pairs of oscillators are, on average, smaller than the mutual lags between unconnected pairs. Here we test if we can use this property to infer the underlying physical connectivity of the R\"ossler or of the Kuramoto oscillators, i.e., to reconstruct the network by classifying the links in two categories, existing and non-existing. 

To do this we first calculate, for each coupling strength, the set of $N(N-1)/2$ similarity values, $S_{ij}=\rho_{ij}$ or $S_{ij}=PLV_{ij}$, and the corresponding set of lags, $\tau_{ij}$, that maximize $\rho_{ij}(\tau)$ or $PLV_{ij}(\tau)$. Then, for each coupling strength we define two thresholds, one for the lags, $\tau_{th}$, and one for the similarity values, $S_{th}$, and use the following criteria to classify a link as existing or not-existing:
\begin{enumerate}
\item SIM: the link between $i$ and $j$ exists if $S_{ij}>S_{th}$, else, it does not exist.
\item AND: the link between $i$ and $j$ exists if  $\tau_{ij}<\tau_{th}$ and $S_{ij}>S_{th}$, else, it does not exist.
\item OR: the link between $i$ and $j$ exists if  $\tau_{ij}<\tau_{th}$ or $S_{ij}>S_{th}$, else, it does not exist.
\end{enumerate}

\subsection{Diagnostic ability of the classification criteria}

To quantify the efficiency of these criteria for uncovering the underlying network structure (i.e., the existing, $A_{ij}=1$, and the non-existing , $A_{ij}=0$, links) we calculate, for each criteria and for each value of the coupling, the area under the receiver operating characteristic (ROC) curve \cite{Fawcett2006}. The ROC curve is obtained by varying the detection threshold and calculating the
\begin{enumerate}
\item True positives (TP):  number of links that are correctly detected as existing.
\item True negatives (TN): number of links that are correctly detected as non-existing.
\item  False positives (FP): number of links that are incorrectly detected as existing.
\item  False negatives (FN): number of links that are incorrectly detected as non-existing.
\end{enumerate}

Then, the true positive rate (TPR, also known as {\it {recall}}) is TP/(TP+FN)=TP/(\# of existing links), and the false positive rate (FPR) is FP/(TN+FP)=FP/(\# of non existing links). A ROC curve is obtained by ploting the TPR vs. FPR. 

The area under the ROC curve (AUC) is a measure of the goodness of a binary classifier: while random guessing gives a diagonal line, a perfect classifier has one (or more) thresholds that perfectly separate the existing and the non-existing links. In this situation, AUC=1. 

While the AUC is routinely used to quantify the performance of a binary classifier, in class imbalance scenarios (for example, when there are a lot of patients without a decease) it has been shown that the precision-recall curve is more informative because it does not depend on the number of true negatives \cite{precision}. The {\it {precision}} is the ratio of correct positive detections over all positive detections, TP/(TP+FP), and the precision-recall curve is obtained by plotting the precision vs. the recall (i.e., the TPR).

Because our networks are sparce (only about 10\% of the possible links exist) we also use as a performance measure the {\it {average precision}} that  is the area under the precision-recall curve.


\section{Results}
\begin{figure}[h!] 
\begin{center}
\includegraphics[width=.7\textwidth]{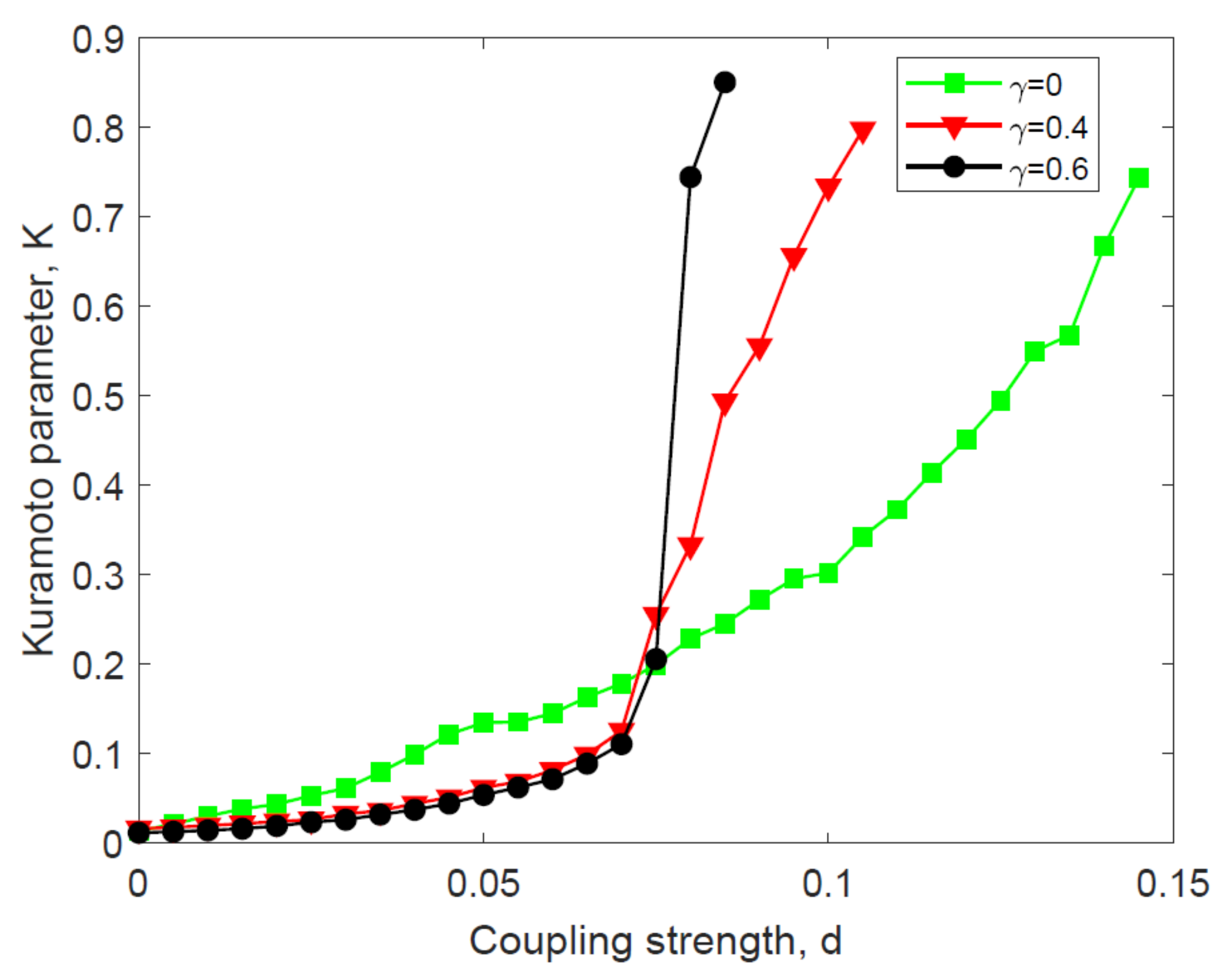}
\end{center}
\caption{Kuramoto order parameter as a function of the coupling strength, for different values of the frequency gap, $\gamma$. We note that as $\gamma$ increases, the synchronization transition becomes more abrupt.}
\label{fig:KOs}
\end{figure}

We begin by characterizing the synchronization transition as a function of the frequency gap, $\gamma$. Figure \ref{fig:KOs} displays the order parameter, Eq.~\ref{eq:K}, when $\gamma=0$, 0.4 and 0.6.  For $\gamma$=0 the network is constructed  without restrictions, i.e., the links are distributed randomly and therefore, the synchronization transition is second order.  For $\gamma$=0.4 the transition to synchronization is more abrupt and it is explosive for $\gamma$=0.6.

\begin{figure}[h!] 
\begin{center}
\includegraphics[width=.9\textwidth]{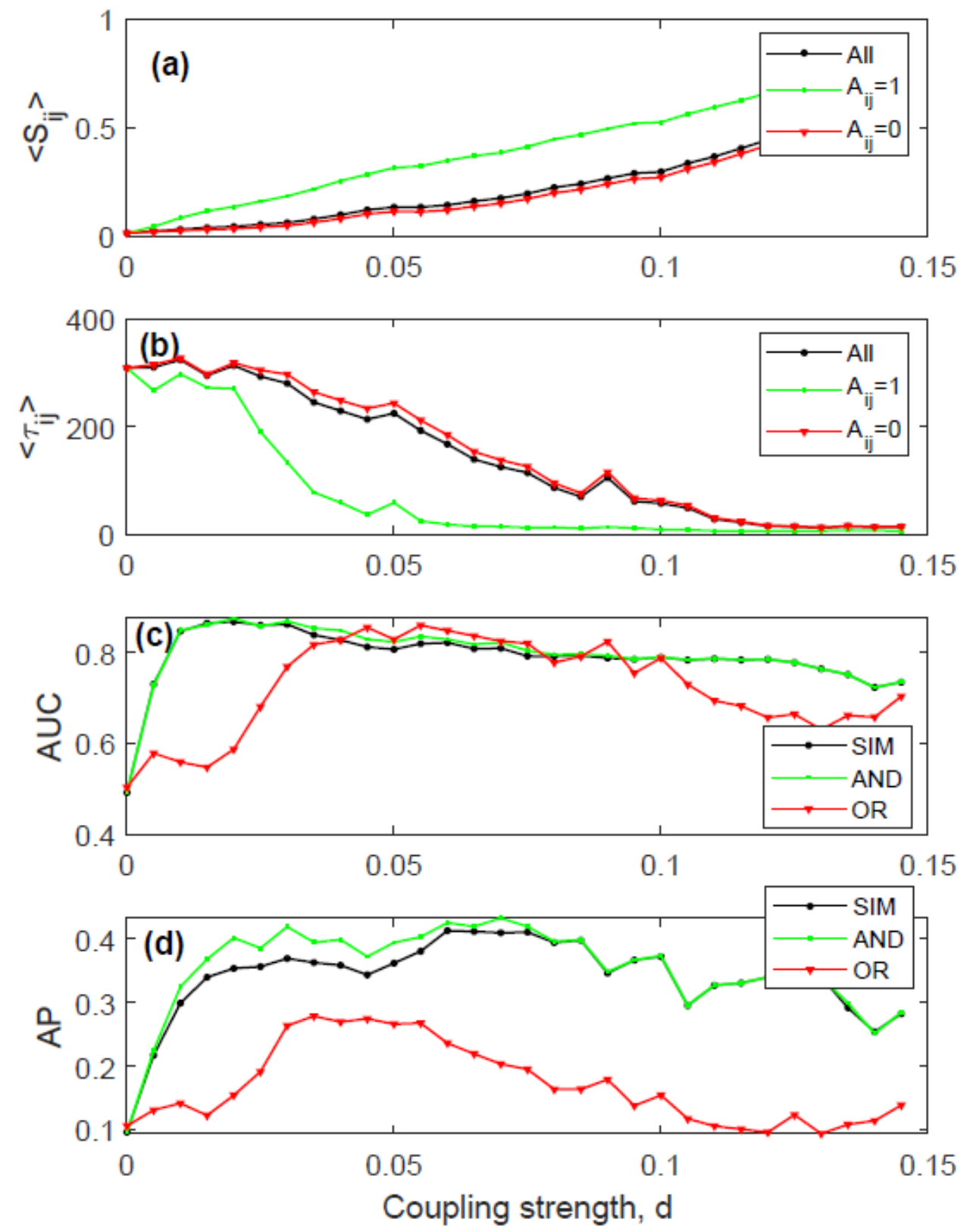}
\end{center}
\caption{Analysis of the Kuramoto oscillators when the frustation parameter is $\gamma=0$. The panels displays the average similarity value (a), the average lag time (b), the area under the ROC curve (c) and the average precision (d).  50 oscillators are simulated with 125 random mutual links (1100 links do not exist). 
The similarity between pairs of oscillators, $S_{ij}$, is measured with the Pearson coefficient, Eq.(\ref{eq:pearson}). The length of the time series is $T=3000$ datapoints and the we search for the maximum value in the interval [0,$\tau_{max}=600$].}
\label{fig:cc_caso1}
\end{figure}

\begin{figure}[h!] 
\begin{center}
\includegraphics[width=.9\textwidth]{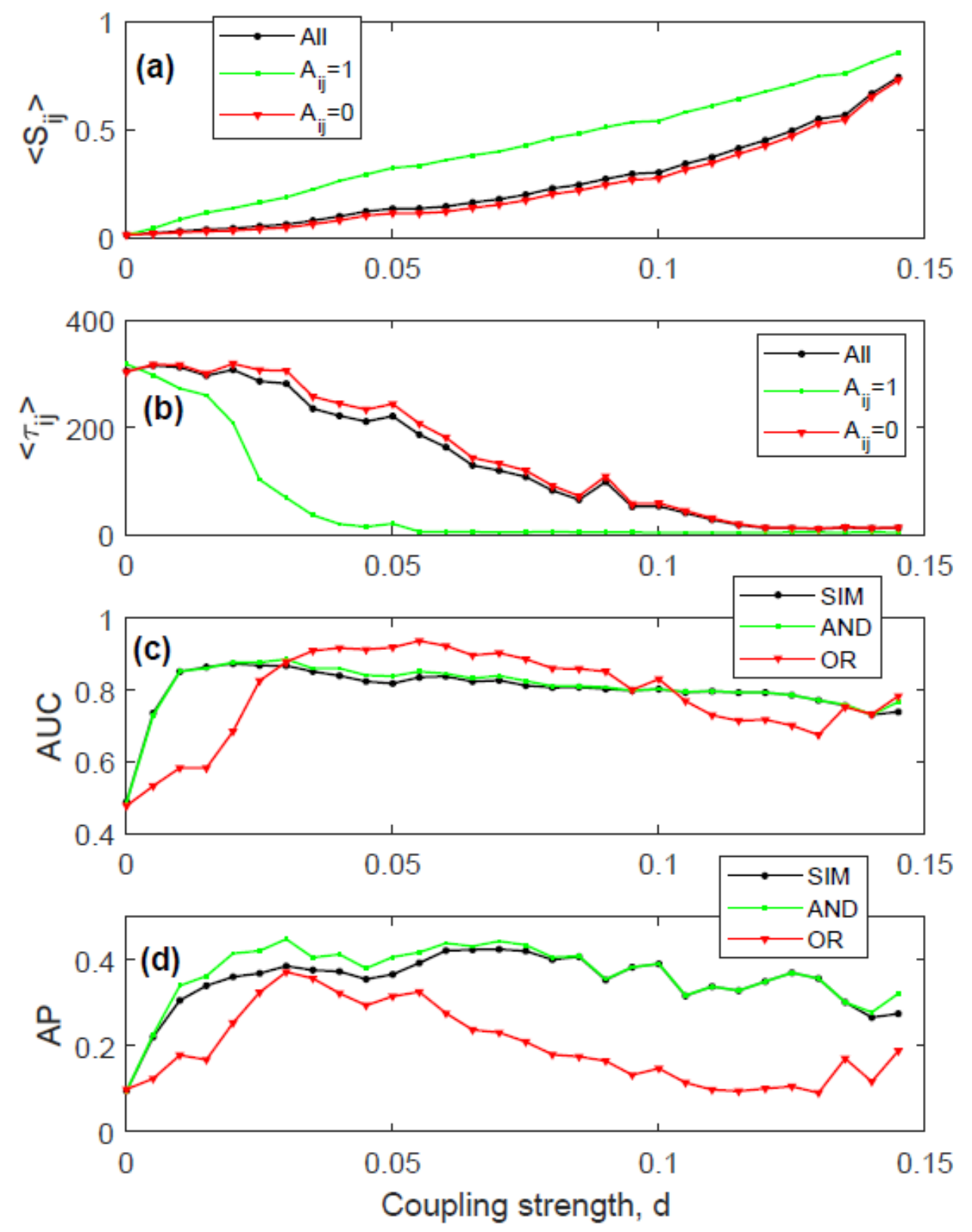}
\end{center}
\caption{As Fig.\ref{fig:cc_caso1} but the similarity between pairs of oscillators is measured with the phase locking value, Eq. (\ref{eq:PLV}).}
\label{fig:plv_caso1}
\end{figure}

Figures~\ref{fig:cc_caso1} and~\ref{fig:plv_caso1} present the analysis of the Kuramoto network when the frequency gap is $\gamma$=0, using the lagged Pearson coefficient (Fig. \ref{fig:cc_caso1}) and the PLV (Fig. \ref{fig:plv_caso1}) respectively. For each procedure, panels (a) displays the correspondent average similarity value $\left<S_{ij}\right>$ (where $S_{ij}=\rho_{ij}$ in Fig.  \ref{fig:cc_caso1} and $S_{ij}=PLV_{ij}$ in Fig. \ref{fig:plv_caso1}), and panel (b) displays $\left<\tau_{ij}\right>$, averaged over all the pairs of oscillators and over time. We note that in both cases (using $\rho_{ij}$ or $PLV_{ij}$), when the $i-j$ link exists $S_{ij}$ tends to be larger than when it does not exist. We also note that, as the coupling strength increases, on average, for the existing links $\tau_{ij}$ starts decreasing much earlier than for the non-existing links. We note that the difference is clear even if the order parameter, $K$, and the average similarity value, $\left<S_{ij}\right>$, don't show any sign of the approaching  transition.  

Regarding the use of the lag-time information for inferring the network structure, the panels (c) and (d) (that display the area under the ROC curve and the average precision, respectively) show that the classification criteria SIM and AND tend to have similar performance, while OR is lower. 

\begin{figure}[h!] 
\begin{center}
\includegraphics[width=.9\textwidth]{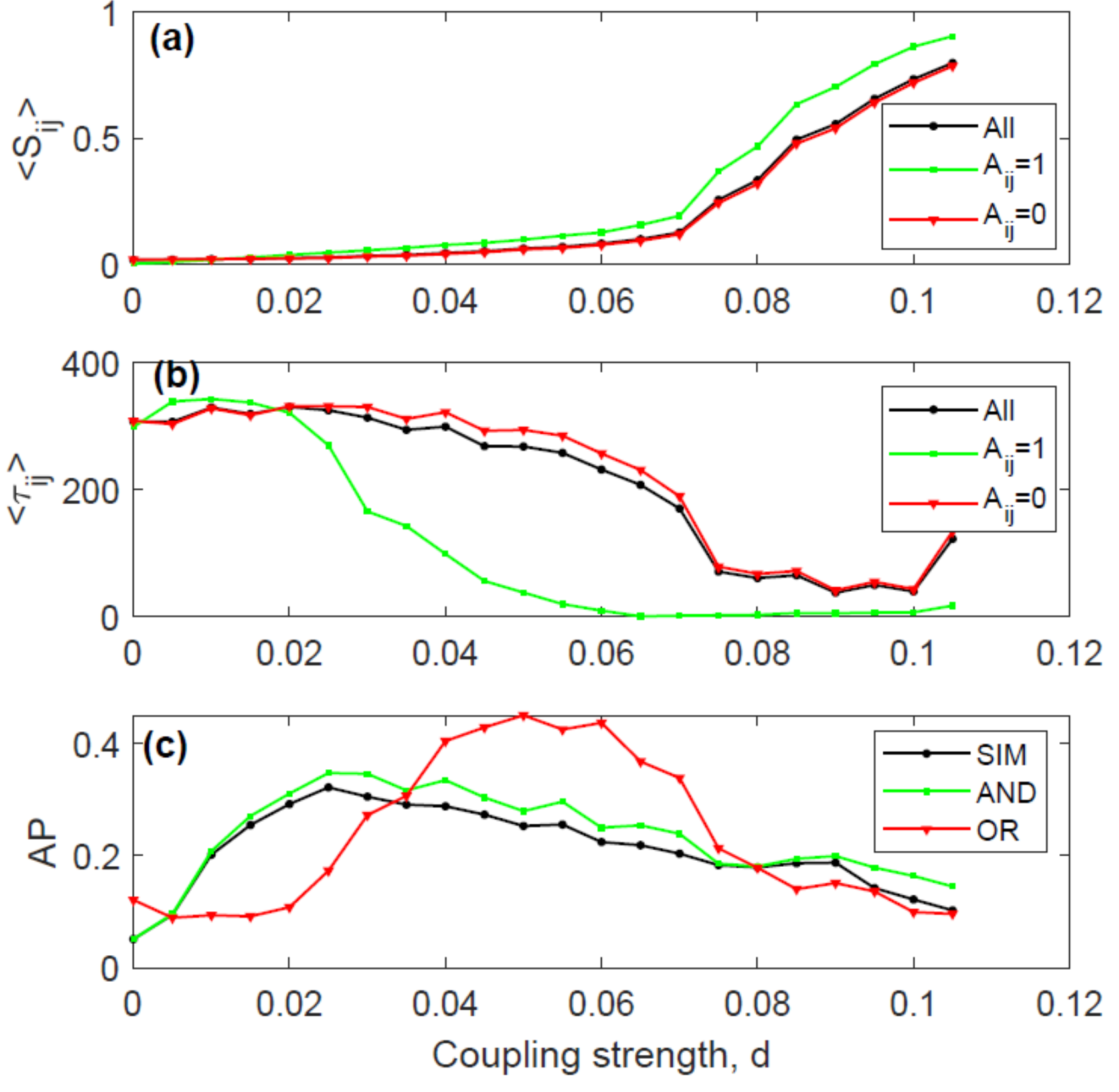}
\end{center}
\caption{Analysis of the Kuramoto oscillators when the frustation parameter is $\gamma=0.4$. (a) Average similarity value (the phase locking value), (b) average lag time, and (c) average precision.}
\label{fig:plv_caso4}
\end{figure}

\begin{figure}[h!] 
\begin{center}
\includegraphics[width=.9\textwidth]{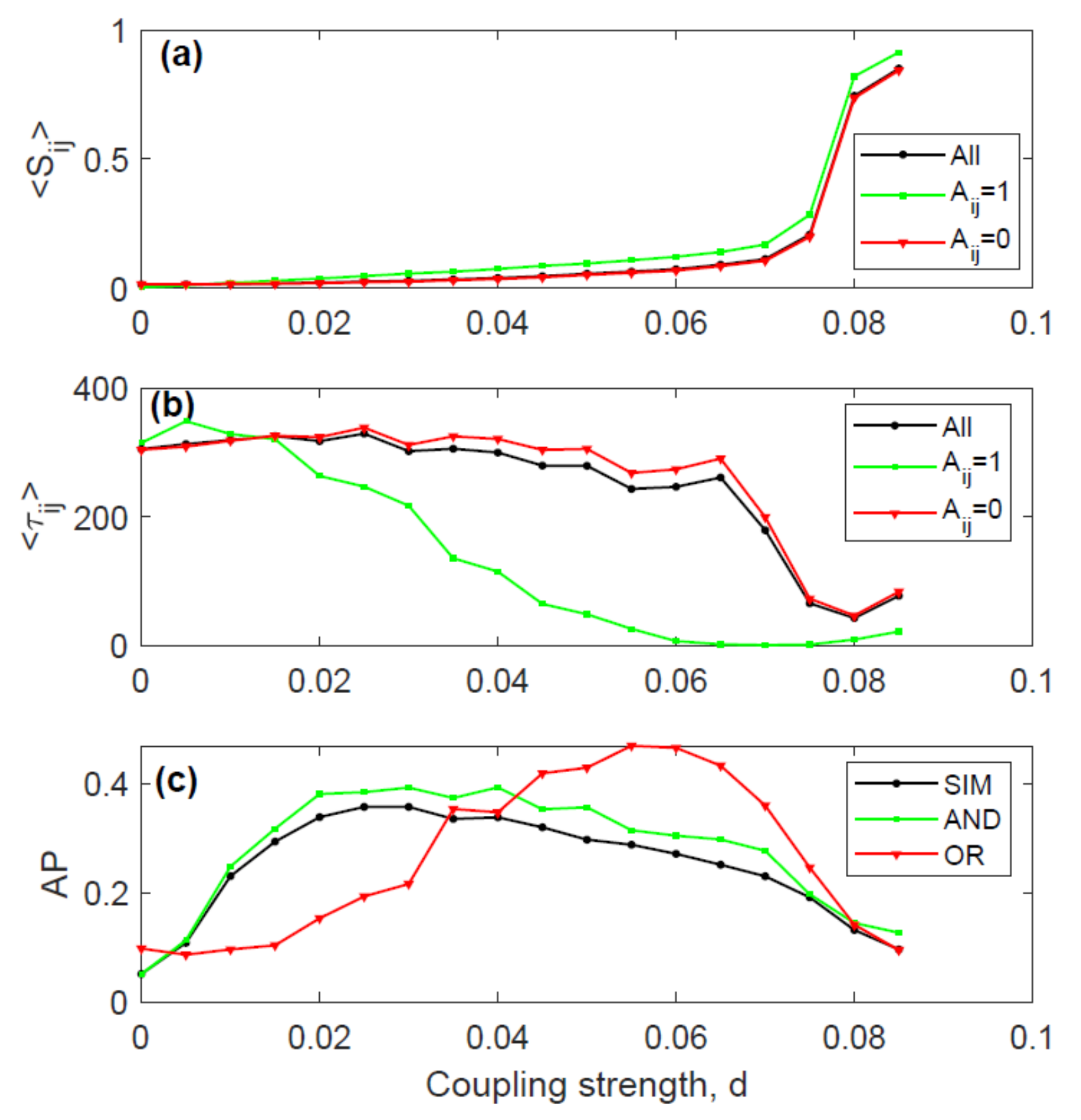}
\end{center}
\caption{As Fig.\ref{fig:plv_caso4} but $\gamma=0.6$.}
\label{fig:plv_caso3}
\end{figure}

In Figs.~\ref{fig:plv_caso4} and ~\ref{fig:plv_caso3} we analyze the role of $\gamma$. We use as similarity measure the PLV and quantify the classification performance with the average precision.  
The behavior when the transition to synchronization is more abrupt (Fig.~\ref{fig:plv_caso4} for $\gamma$=0.4) or even explosive (Fig.~\ref{fig:plv_caso3} for $\gamma$=0.6) is consistent with that found for $\gamma=0$: the difference between the average similarity values of the existing and the non-existing links is small, but the average lags of the existing and the non-existing links are well separated. Therefore, we can consider that the decrease of the mean value of the lag-times of the existing links provides a robust early indication of the transition to synchronization. Regarding network inference, the average precision indicates that, depending on the coupling strength, the delay times can be informative for the detection of the existing links.

\begin{figure}[h!] 
\begin{center}
\includegraphics[width=.9\textwidth]{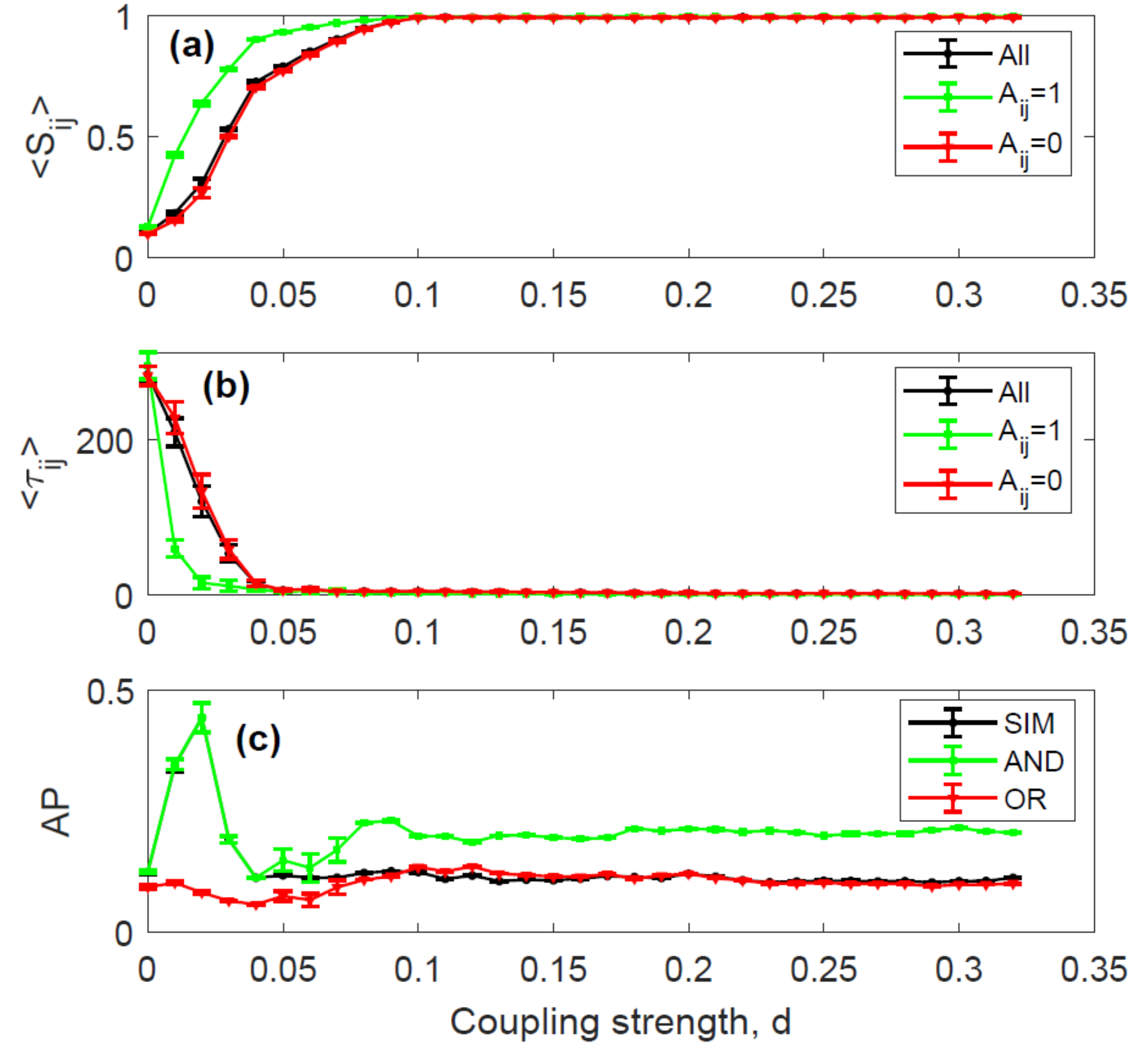}
\end{center}
\caption{Analysis of the experimental dataset recorded from 28 chaotic electronic circuits coupled with 42 mutual links (336 links do not exist). Average similarity value computed using lagged cross-correlation (a), average lag time (b), area under the ROC curve (c) and average precision (d). The error bars represent the standard deviation of the values obtained from 5 non-overlapping windows. $L=5000$, $\tau_{max}=1000$.}
\label{fig:1_28osc}
\end{figure}

Figure~\ref{fig:1_28osc} displays the results of the analysis of the experimental data recorded from the electronic R\"ossler oscillators, and confirms the results obtained with the Kuramoto oscillators.
Here the similarity measure is the lagged cross-correlation method and the coupling topology is purely random, and therefore the experimental system reaches synchronization in a continuous second order transition. As in the numerical system, we note that $S_{ij}$  ($\tau_{ij}$) tends to be larger (smaller) when the link exists than when it does not exist. In Fig.~\ref{fig:1_28osc} we see that the system reaches a high synchronization level with $\left<S_{ij}\right>\sim 1$. We also note that for large enough coupling the lag information with the AND criterion is indeed useful for improving the inference of the network in comparison to the SIM or OR criteria.

\section{Conclusions}
We have analyzed numerical and experimental datasets representing the dynamics of networked oscillators: 50 Kuramoto phase oscillators and  28 R\"ossler electronic chaotic circuits respectively. The goal of our study was to determine if the mutual lags between the oscillators contain relevant information for inferring the structural network of connections and/or the global behavior of the system.  For the Kuramoto oscillators we have performed the analysis using two methods of computing the node-to-node similarity: a cross-correlation analysis of the ``observable'' $\cos[\phi(t)]$, and the lagged phase locking value, PLV. For the experimental data, the analysis was performed over the raw time series of the observed dynamical variable using lagged cross-correlation. In both cases we found that the values of the lag-times between oscillators, as the coupling strength increases, carry information that can be used as an early indicator of the transition to synchrony, and/or for inferring the oscillators' connectivity.  The results obtained are fully consistent for both methods of computing the oscillators' similarity. The calculation of the PLV has the disadvantage that it can be used only for data sets for which meaningful phases can be defined, while the cross-correlation can be used for any data set.
  
Future work will aim to test further the robustness of these observations over larger networks of elements coupled with directed and heterogeneous links. Future work will also aim to highlight any difference in the  performance of the network inference that can depend on the method of data analysis used to determine the similarity values and lag times.

\section*{Acknowledgments}

This work was supported in part by the Ibersinc network funded by Spanish MINECO grant FIS2015-71929-REDT.
C.M. also acknowledges partial support from Spanish Ministerio de Ciencia, Innovación y Universidades grant PGC2018-099443-B-I00 and ICREA ACADEMIA, Generalitat de Catalunya.  I. L. acknowledges support from the Spanish MINECO  through grant FIS2017-84151-P. 


\bibliography{test}

\end{document}